\documentclass[aps,preprint]{revtex4}%
\usepackage{amsfonts}
\usepackage{amsmath}
\usepackage{amssymb}
\usepackage{graphicx}%
\providecommand{\U}[1]{\protect\rule{.1in}{.1in}}

\begin{document}
\author{}
\title{Vector particles tunneling from BTZ black holes}
\author{Ge-Rui Chen}
\email{chengerui@emails.bjut.edu.cn}
\author{Shiwei Zhou}
\author{Yong-Chang Huang}
\affiliation{Institute of Theoretical Physics, Beijing University of
Technology, Beijing, 100124, China}

\begin{abstract}
In this paper we investigate vector particles' Hawking radiation
from a BTZ black hole. By applying the WKB approximation and the
Hamilton-Jacobi Ansatz to the Proca equation, we obtain the
tunneling spectrum of vector particles. The expected Hawking
temperature is recovered.

Keywords: {\ Vector particles tunneling, BTZ black hole, Proca
equation}
\end{abstract}

\pacs{97.60.Lf; 04.70.Dy}
\maketitle

Since Stephen Hawking discovered that black holes radiate black body
spectrum\cite{sk1,sk2}, there are several different methods to
derive Hawking radiation to confirm this astounding discovery. In
recent years, semi-classical methods of modeling Hawking radiation
as a tunneling process have garnered lots of interest. This
semi-classical methods which model Hawking radiation as a tunneling
process use WKB approximation to calculate the imaginary part of the
action for the classically forbidden trajectory across the horizon.
The tunneling probability is given by $\Gamma\propto\exp(-2Im I)$.

The first black hole tunneling method--Null Geodesic Method was
proposed by Parikh and Wilczek\cite{mkp1} which followed from the
work of Kraus and Wilczek\cite{mkp2}. They found the only part of
the action that contributes an imaginary term is $Im
I=\int_{r_{in}}^{r_{out}}p_r dr$, where $p_r$ is the momentum of the
emitted null s-wave. One can use Hamilton's equation and knowledge
of the null geodesics to calculate the imaginary part of the action.
Later on, Refs.\cite{jwc,jz} extended this method to the charged
particles. The other method is the Hamilton-Jacobi Ansatz used by
Refs.\cite{ma,rk}, which is an extension of the complex path
analysis of Padmanabhan et al\cite{tp}. This method applies the WKB
approximation to the Klein-Gordon equation, and the lowest order is
the Hamilton-Jacobi equation. Then according to the symmetry of the
metric, one can pick an appropriate ansatz for the action and put it
into the Hamilton-Jacobi equation to solve. In 2008, R. Kerner and
R.B. Mann\cite{rk2,rk3} applied the WKB approximation to the Dirac
Equation to calculate Dirac particles' Hawking radiation.

Recently, S. I. Kruglov\cite{sik2} investigated black hole radiation
of vector particles in (1+1) dimensions by using the Hamilton-Jacobi
method to the Proca equation. Vector particles(e.g. $Z,W^{\pm}$)
play important role in Standard Model, so it is interesting to study
the Hawking radiation of vector particles. We extend
Ref.\cite{sik2}'s method to investigate vector particles tunneling
from a BTZ black hole and obtain the radiation spectrum for vector
particles. Our results show that vector particles radiate with the
standard Hawking temperature. Like Refs.\cite{rk2,rk3}, we assume
that the change of black hole angular momentum due to the spin of
the emitted particle is negligible. This is a good approximation for
the black hole with mass much larger than the Planck mass.

The BTZ back hole is the solution of the standard Einstein-Maxwell
equation in $(2+1)$ dimensional spacetime with a negative cosmology
constant. For simplicity we will ignore the coupling to the Maxwell
field. The action is
\begin{eqnarray}
I=\frac{1}{2\pi}\int\sqrt{-g}[R+2l^{-2}]dx^2dt+B,\label{21}
\end{eqnarray}
where $B$ is a surface term and the radius $l$ is related to the
cosmological constant by $\Lambda=-l^{-2}$. The equations of motion
derived from (\ref{21}) are solved by the black hole field
\begin{eqnarray}
ds^2&=&-(-M+\frac{r^2}{l^2}+\frac{J^2}{4r^2})dt^2+(-M+\frac{r^2}{l^2}+\frac{J^2}{4r^2})^{-1}dr^2+r^2(d\phi-\frac{J}{2r^2}dt)^2\nonumber\\
&=&-(-M+\frac{r^2}{l^2})dt^2+(-M+\frac{r^2}{l^2}+\frac{J^2}{4r^2})^{-1}dr^2+r^2d\phi^2-Jdtd\phi\label{22}
\end{eqnarray}
with $-\infty<t<\infty, 0<r<\infty$ and $0\leq\phi<2\pi$. The two
constants of integration $M$ and $J$ appearing in (2) are the
conserved charges associated with asymptotic invariance under time
displacements (mass) and rotational invariance (angular momentum),
respectively. The metric (\ref{22}) has two horizons
\begin{eqnarray}
r_\pm=lM^\frac{1}{2}\left[\frac{1}{2}\left(1\pm\sqrt{1-(\frac{J}{Ml})^2}\right)\right]^\frac{1}{2},
\end{eqnarray}
where $r_+$ is the black hole horizon. In order for the horizon to
exist one must have
\begin{eqnarray}
M>0, |J|\leq Ml.
\end{eqnarray}
In the extreme case $|J| = Ml$, both roots of $g_{11}^{-1}$
coincide. Let us define the following notations for convenience
\begin{eqnarray}
A(r)=-M+\frac{r^2}{l^2}, \
B(r)=-M+\frac{r^2}{l^2}+\frac{J^2}{4r^2},\  C=r^2, \ 2D=-J,
\end{eqnarray}
so the metric (\ref{22}) becomes
\begin{eqnarray}
ds^2=-A(r)dt^2+\frac{1}{B(r)}dr^2+Cd\phi^2+2Ddtd\phi.
\end{eqnarray}
The determinant is
\begin{eqnarray}
\sqrt{-g}=\sqrt{\frac{AC+D^2}{B}},
\end{eqnarray}
and the inverse metric is
\begin{eqnarray}
 g^{\mu \nu}=-\frac{1}{AC+D^2}
\left(\begin{array}{ccc}
   C & 0 & -D   \\
   0 & -B(AC+D^2) & 0   \\
   -D & 0 & -A   \\
 \end{array}\right).
\end{eqnarray}

The Proca equation for vector particles is\cite{sik2}
\begin{eqnarray}
D_\mu\psi^{\nu\mu}+\frac{m^2}{\hbar^2}\psi^{\nu}=0,
\end{eqnarray}
\begin{eqnarray}
\psi_{\nu\mu}=D_\nu\psi_\mu-D_\mu\psi_\nu=\partial_\nu\psi_\mu-\partial_\mu\psi_\nu,
\end{eqnarray}
where $D_\mu$ are covariant derivatives, and
$\psi_\nu=(\psi_0,\psi_1,\psi_2)$.  From the definition,
$\psi^{\nu\mu}$ is an anti-symmetrical tensor, so using the equation
\begin{eqnarray}
D_\mu\psi^{\nu\mu}=\frac{1}{\sqrt{-g}}\partial_\mu(\sqrt{-g}\psi^{\nu\mu}),
\end{eqnarray}
the Proca equation becomes
\begin{eqnarray}
\frac{1}{\sqrt{-g}}\partial_\mu(\sqrt{-g}\psi^{\nu\mu})+\frac{m^2}{\hbar^2}\psi^\nu=0.
\end{eqnarray}
From the metric and the following relationship
\begin{eqnarray}
\psi^0&=&\psi_\mu g^{\mu0}=\frac{-C}{AC+D^2}\psi_0+\frac{D}{AC+D^2}\psi_2,\nonumber\\
\psi^1&=&B\psi_1, \ \psi^2=\frac{D}{AC+D^2}\psi_0+\frac{A}{AC+D^2}\psi_2,\nonumber\\
\psi^{01}&=&\psi_{\mu\nu}g^{\mu0}g^{\nu1}=-\frac{CB}{AC+D^2}\psi_{01}-\frac{DB}{AC+D^2}\psi_{12},\nonumber\\
\psi^{02}&=&-\frac{1}{AC+D^2}\psi_{02}, \
\psi^{12}=-\frac{BD}{AC+D^2}\psi_{01}+\frac{BA}{AC+D^2}\psi_{12},
\end{eqnarray}
we obtain the Proca equations in the following explicit form
\begin{eqnarray}
\frac{1}{\sqrt{-g}}\{\partial_r[\sqrt{-g}(\frac{-CB}{AC+D^2}(\partial_t\psi_1-\partial_r\psi_0)-\frac{DB}{AC+D^2}(\partial_r\psi_2-\partial_\phi\psi_1))]\nonumber\\
+\partial_\phi[\sqrt{-g}(\frac{-1}{AC+D^2}(\partial_t\psi_2-\partial_\phi\psi_0))]\}+\frac{m^2}{\hbar^2}(\frac{-C}{AC+D^2}\psi_0+\frac{D}{AC+D^2}\psi_2)=0,\nonumber\\
\frac{1}{\sqrt{-g}}\{\partial_t[\sqrt{-g}(\frac{CB}{AC+D^2}(\partial_t\psi_1-\partial_r\psi_0)+\frac{DB}{AC+D^2}(\partial_r\psi_2-\partial_\phi\psi_1))]\nonumber\\
+\partial_\phi[\sqrt{-g}(\frac{BD}{AC+D^2}(\partial_r\psi_0-\partial_t\psi_1)+\frac{AB}{AC+D^2}(\partial_r\psi_2-\partial_\phi\psi_1))]\}+\frac{m^2}{\hbar^2}B\psi_1=0,\nonumber\\
\frac{1}{\sqrt{-g}}\{\partial_t[\sqrt{-g}(\frac{1}{AC+D^2}(\partial_t\psi_2-\partial_\phi\psi_0))]
+\partial_r[\sqrt{-g}(\frac{BD}{AC+D^2}(\partial_t\psi_1-\partial_r\psi_0)\nonumber\\
-\frac{AB}{AC+D^2}(\partial_r\psi_2-\partial_\phi\psi_1))]\}+\frac{m^2}{\hbar^2}(\frac{D}{AC+D^2}\psi_0+\frac{A}{AC+D^2}\psi_2)=0.\label{13}
\end{eqnarray}

According to the WKB approximation, the solutions are in the form
\begin{eqnarray}
\psi_\nu=(c_0,c_1,c_2)\exp(\frac{i}{\hbar}S(t,r,\phi)),\label{11}
\end{eqnarray}
where
\begin{eqnarray}
S(t,r,\phi)=S_0(t,r,\phi)+\hbar
S_1(t,r,\phi)+\hbar^2S_2(t,r,\phi)+\cdots\label{12}
\end{eqnarray}
Putting Eqs.(\ref{11},\ref{12}) into Eqs.(\ref{13}), the equations
to the leading order in $\hbar$ are
\begin{eqnarray}
&&\frac{CB}{AC+D^2}[c_1(\partial_t s_0)(\partial_r
s_0)-c_0(\partial_rs_0)^2]+\frac{BD}{AC+D^2}[c_2(\partial_r
s_0)^2-c_1(\partial_\phi
s_0)(\partial_rs_0)]\nonumber\\
&&+\frac{1}{AC+D^2}[c_2(\partial_ts_0)(\partial_\phi
s_0)-c_0(\partial_\phi
s_0)^2]-\frac{C}{AC+D^2}m^2c_0+\frac{D}{AC+D^2}m^2c_2=0,\nonumber\\
&&\frac{CB}{AC+D^2}[-c_1(\partial_t
s_0)^2+c_0(\partial_rs_0)(\partial_t
s_0)]+\frac{BD}{AC+D^2}[-c_2(\partial_r s_0)(\partial_t
s_0)+c_1(\partial_\phi
s_0)(\partial_ts_0)]\nonumber\\
&&+\frac{BD}{AC+D^2}[-c_0(\partial_rs_0)(\partial_\phi
s_0)+c_1(\partial_ts_0)(\partial_\phi
s_0)]\nonumber\\
&&+\frac{AB}{AC+D^2}[-c_2(\partial_rs_0)(\partial_\phi
s_0)+c_1(\partial_\phi s_0)^2]+m^2Bc_1=0,\nonumber\\
&&\frac{1}{AC+D^2}[-c_2(\partial_t s_0)^2+c_0(\partial_\phi
s_0)(\partial_t s_0)]+ \frac{BD}{AC+D^2}[-c_1(\partial_t
s_0)(\partial_rs_0)+c_0(\partial_rs_0)^2]\nonumber\\
&&-\frac{AB}{AC+D^2}[-c_2(\partial_rs_0)^2+c_1(\partial_\phi
s_0)(\partial_r
s_0)]+m^2\frac{D}{AC+D^2}c_0+m^2\frac{A}{AC+D^2}c_2=0.\label{23}
\end{eqnarray}
There exists a solution of the form
\begin{eqnarray}
S_0=-Et+j\phi+W(r)+K,\label{14}
\end{eqnarray}
where $E=-\partial_tS_0$ and $j$ are the energy and angular momentum
of the emitting particles respectively. $K$ is a complex constant.
Inserting Eq.(\ref{14}) into Eqs.(\ref{23}) we obtain the matrix
equation
\begin{eqnarray}
\Lambda(c_0,c_1,c_2,)^T=0,\label{16}
\end{eqnarray}
where $\Lambda$ is a $3\times3$ matrix, and its components are
expressed as
\begin{eqnarray}
\Lambda_{11}&=&-CB(W^\prime)^2-j^2-Cm^2,\
\Lambda_{12}=-CBEW^\prime-BDjW^\prime,\nonumber\\
\Lambda_{13}&=&BD(W^\prime)^2-Ej+Dm^2,\
\Lambda_{21}=-CBEW^\prime-BDW^\prime j,\nonumber\\
\Lambda_{22}&=&-CBE^2-2BDEj+ABj^2+B(AC+D^2)m^2,\
\Lambda_{23}=BDEW^\prime-ABjW^\prime,\nonumber\\
\Lambda_{31}&=&-Ej+BD(W^\prime)^2+m^2D,\
\Lambda_{32}=BDEW^\prime-BAjW^\prime,\nonumber\\
\Lambda_{33}&=&-E^2+BA(W^\prime)^2+m^2A,
\end{eqnarray}
where $W^\prime=\partial_rS_0$ and $j=\partial_\phi S_0$.

Homogeneous system of linear equations (\ref{16}) possesses
nontrivial solution if the determinant of the matrix $\Lambda$
equals to zero, that is, det$\Lambda=0$. After the calculation, we
have
\begin{eqnarray}
-Bm^2\{-2DEj+Aj^2+D^2[m^2+B(W^\prime)^2]+C[-E^2+A(m^2+B(W^\prime)^2)]\}^2=0.
\end{eqnarray}
We obtain immediately
\begin{eqnarray}
(W^\prime)^2&=&\frac{CE^2+2DEj-Aj^2-m^2(D^2+CA)}{B(D^2+CA)}\nonumber\\
&=&\frac{(E-\frac{J}{2r^2}j)^2-(-M+\frac{r^2}{l^2}+\frac{J^2}{4r^2})(m^2+\frac{j^2}{r^2})}{(-M+\frac{r^2}{l^2}+\frac{J^2}{4r^2})^2},
\end{eqnarray}
and
\begin{eqnarray}
W_{\pm}=\pm\int\frac{\sqrt{(E-\frac{J}{2r^2}j)^2-(-M+\frac{r^2}{l^2}+\frac{J^2}{4r^2})(m^2+\frac{j^2}{r^2})}}{-M+\frac{r^2}{l^2}+\frac{J^2}{4r^2}}dr.
\end{eqnarray}
One solution $W_+$ corresponds to vector particles moving away from
the black hole, and the other $W_-$ corresponds to particles moving
toward the black hole. Based on the discussion of Ref.\cite{rk2},
the probability of tunneling particles is
\begin{eqnarray}
\Gamma\propto\exp(-\frac{4}{\hbar}ImW_+).
\end{eqnarray}
Integrating around the pole at the horizon $r_+$, we obtain
\begin{eqnarray}
ImW_+=\frac{\pi}{2\kappa}(E-\Omega_+j),
\end{eqnarray}
where $\kappa=\frac{r_+^2-r_-^2}{l^2r_+}$ is the surface gravity of
outer event horizon and
$\Omega_+\equiv-\frac{g_{02}}{g_{22}}|_{r_+}=\frac{J}{2r_+^2}$ is
the angular velocity of the outer event horizon. So the tunneling
probability is
\begin{eqnarray}
\Gamma\propto\exp[-\frac{2 \pi}{\kappa}(E-\Omega_+j)],
\end{eqnarray}
and the Hawking temperature is
\begin{eqnarray}
T_H=\frac{\kappa}{2\pi}=\frac{r_+^2-r_-^2}{2\pi l^2r_+}.
\end{eqnarray}
We get the Hawking temperature of vector particles from BTZ black
holes. The result is the same as Dirac particles tunneling from BTZ
black holes\cite{rl}.

In summary, we investigate vector particles' Hawking radiation from
a BTZ black hole. By applying the WKB approximation and the
Hamilton-Jacobi Ansatz to the Proca equation, we obtain the
tunneling spectrum of vector particles and the expected Hawking
temperature, which provides further evidence for the universality of
back radiation. Compared with other methods to calculate Hawking
radiation, Hamilton-Jacobi method has a clear physical picture, is
straightforward to compute, and easily generalize to different kinds
of particles.

\begin{acknowledgements}
This work is supported by National Natural Science Foundation of
China (No.11275017 and No.11173028).
\end{acknowledgements}

\end{document}